\documentclass[10pt,journal,a4paper,final,oneside,twocolumn]{IEEEtran}
\usepackage{amsmath,amsfonts}
\usepackage{array}
\usepackage[caption=false,font=normalsize,labelfont=sf,textfont=sf]{subfig}
\usepackage{textcomp}
\usepackage{stfloats}
\usepackage{url}
\usepackage{verbatim}
\usepackage{graphicx}

\usepackage{booktabs}
\usepackage{multirow}
\usepackage{makecell}
\usepackage{cite}
\usepackage{enumitem}

\usepackage[ruled,noend,algosection,noresetcount]{algorithm2e}
\begin{document}

\title{An End-to-End Neural Network Transceiver Design for OFDM System with FPGA-Accelerated Implementation}

\author{
    Yi Luo,~
    Luping Xiang,~\IEEEmembership{Senior Member, IEEE,} 
    Cheng Luo,~\IEEEmembership{Graduate Student Member, IEEE,}
    Kun Yang,~\IEEEmembership{Fellow,~IEEE,}
    Shida Zhong,~\IEEEmembership{Member, IEEE,}
    and Jienan Chen,~\IEEEmembership{Senior Member, IEEE}
    
\thanks{Yi Luo is with the School of Information and Communication Engineering, University of Electronic Science and Technology of China, Chengdu 611731, China (e-mail: 202421011115@std.uestc.edu.cn).}

\thanks{Luping Xiang, Cheng Luo, and Kun Yang are with the State Key Laboratory of Novel Software Technology, Nanjing University, Nanjing, 210008, China, Institute of Intelligent Networks and Communications (NINE), Nanjing University (Suzhou Campus), Suzhou, 215163, China (e-mail: luping.xiang@nju.edu.cn, joffreylc1997@163.com, kunyang@nju.edu.cn).}

\thanks{Shida Zhong is with the College of Electronics and Information Engineering, Shenzhen University, Shenzhen 518060, China (e-mail: shida.zhong@szu.edu.cn).}

\thanks{Jienan Chen is with the National Key Laboratory of Science and Technology on Communications of the University of Electronic Science and Technology of China, Chengdu 611731, China (e-mail: Jesson.chen@outlook.com).}
 
}

\markboth{}{}

\IEEEpubid{}

\maketitle

\begin{abstract}
The evolution toward sixth-generation (6G) wireless networks demands high-performance transceiver architectures capable of handling complex and dynamic environments. Conventional orthogonal frequency-division multiplexing (OFDM) receivers rely on cascaded discrete Fourier transform (DFT) and demodulation blocks, which are prone to inter-stage error propagation and suboptimal global performance. In this work, we propose two neural network (NN) models DFT-Net and Demodulation-Net (Demod-Net) to jointly replace the IDFT/DFT and demodulation modules in an OFDM transceiver. The models are trained end-to-end (E2E) to minimize bit error rate (BER) while preserving operator equivalence for hybrid deployment. A customized DFT–Demodulation Net Accelerator (DDNA) is further developed to efficiently map the proposed networks onto field-programmable gate array (FPGA) platforms. Leveraging fine-grained pipelining and block matrix operations, DDNA achieves high throughput and flexibility under stringent latency constraints. Experimental results show that the DL-based transceiver consistently outperforms the conventional OFDM system across multiple modulation schemes. With only a modest increase in hardware resource usage, it achieves approximately 1.5 dB BER gain and up to 66\% lower execution time. 
\end{abstract}


\section{Introduction}
\IEEEPARstart{D}{riven} by the rapid proliferation of advanced wireless applications, ranging from smart-device connectivity and the emerging low-altitude economy to massive Internet-of-Things (IoT) scenarios~\cite{ref1}, fifth-generation (5G) networks are continuously evolving toward sixth-generation (6G) systems. High-performance wireless technologies are indispensable for enhancing quality of life and driving socioeconomic advancement~\cite{ref2}. Nevertheless, in realistic environments, the received signal is inevitably impaired by non-ideal factors such as multipath fading, Doppler effects, interference, hardware imperfections, and thermal noise.

Conventional receivers adopt a cascaded physical-layer (PHY) pipeline composed of synchronization, channel estimation, equalization, demapping, and decoding, where each block is optimized sequentially to maximize local performance~\cite{ref3}. This block-wise optimization paradigm, however, suffers from intrinsic limitations: (1) while individual modules may achieve local optimality, the cascaded system is not guaranteed to be globally optimal, and imperfections propagate through successive stages, leading to cumulative errors; and (2) as signal characteristics grow increasingly complex, each module must support richer functionalities and adaptive operations under varying channel conditions, substantially increasing overall system complexity and resource consumption in terms of computation, memory, and power.

In an Orthogonal Frequency Division Multiplexing (OFDM) baseband link, the discrete Fourier transform (DFT) and its inverse (IDFT) form the core mapping between time and frequency domains. Hardware implementations typically utilize the fast Fourier transform (FFT) to achieve efficient computation owing to its reduced arithmetic complexity and hardware efficiency. The inverse FFT (IFFT) can be derived by conjugating or sign-reversing the FFT twiddle factors. Classical FFT algorithms, such as the Cooley–Tukey FFT~\cite{ref4} and the Winograd minimal-multiplication FFT~\cite{ref5}, achieve $O(N\log{N})$ complexity while minimizing multiplication and addition counts. Recent advances exploit algorithmic innovations~\cite{ref6,ref7} and parallel or pipelined architectures~\cite{ref8,ref9} to further reduce latency and memory footprint. To satisfy real-time transformation requirements with varying OFDM subcarrier configurations, several works~\cite{ref9,ref10,ref36} have proposed reconfigurable hardware intellectual property (IP) cores. However, such flexibility incurs increased architectural complexity and greater on-chip resource utilization due to storage of twiddle factors and buffering for data reordering, often resulting in uneven utilization under multi-scenario operation. The Xilinx FFT IP core~\cite{ref11}, for instance, offers multiple FFT architectures that are extensively optimized for hardware deployment.

With the rapid advancement of artificial intelligence, deep learning (DL) has emerged as a promising paradigm for end-to-end (E2E) physical-layer design. Recent studies~\cite{ref3,ref12,ref13,ref14} demonstrate that DL-based E2E PHY architectures can achieve superior robustness and improved adaptability to nonlinear and time-varying impairments. Neural networks (NNs) have been employed to replace specific signal-processing modules, fully reconstruct the receiver, or jointly optimize the entire transceiver chain. In partial substitution, DL models are integrated into individual stages such as modulation classification, channel estimation, equalization, and signal detection~\cite{ref15,ref16,ref17,ref34}, often outperforming conventional algorithms. Other studies explore replacing multiple modules jointly~\cite{ref18,ref19,ref20,ref35}. For example,~\cite{ref18} presents a comprehensive DL framework that simultaneously performs carrier frequency offset compensation, channel estimation, and equalization. DeepWiPHY~\cite{ref19} employs dual jointly trained deep neural networks (DNNs) to replace several receiver blocks and directly generate constellation symbols from received signals. The DCCN framework~\cite{ref20} utilizes deep complex convolutional networks to recover bits from OFDM waveforms without explicit DFT operations, while~\cite{ref35} leverages recurrent neural networks (RNNs) for joint equalization and demodulation to mitigate inter-carrier interference (ICI). Building upon these efforts, this paper seeks to replace key signal-processing modules in OFDM receivers with NNs and jointly optimize overall reception performance. Specifically, we substitute the DFT and demodulation blocks with NN models and benchmark their performance against conventional receivers. Furthermore, unlike prior works, we incorporate an NN-based IDFT at the transmitter and jointly train both ends in an E2E framework. To ensure compatibility with hybrid deployments, the NN model interfaces are designed to align with the standard OFDM processing pipeline.

Practical deployment of DL-based E2E PHY architectures necessitates careful consideration of hardware constraints. The coexistence of extremely high radio-signal sampling rates and limited spectral bandwidth imposes stringent requirements on base-station processing. Therefore, real-time inference at the network edge, subject to strict latency and resource constraints, becomes imperative. However, large NNs with numerous parameters and layers inherently incur high computational and memory demands~\cite{ref21}. The architectures proposed in~\cite{ref20,ref34,ref35} exhibit substantial computational complexity, diverse dataflow characteristics, and temporal dependencies, all of which hinder efficient hardware implementation. Consequently, our work aims to design lightweight, hardware-efficient NN models that balance accuracy, latency, and resource utilization under real-time conditions. Field-programmable gate arrays (FPGAs) are selected as the deployment platform owing to their reconfigurability, low power consumption, and high parallelism, which collectively enable efficient inference~\cite{ref22,ref23,ref24}.
\par
Several studies have mapped DL-based PHY frameworks onto FPGA platforms. For instance,~\cite{ref25,ref26} utilize fully connected neural networks (FCNNs) for signal estimation, mapping the trained models onto Zynq system-on-chip (SoC) devices through hardware–software co-design and fixed-point quantization, achieving performance comparable to that of the linear minimum mean-square error (LMMSE) algorithm. \cite{ref37} applies a CNN to the demodulation of $m$-QAM signals and implements the hardware deployment using Xilinx’s open-source FINN framework. Nevertheless, these designs remain highly specialized, any change in NN configuration necessitates hardware redesign, limiting adaptability in reconfigurable scenarios. Similarly,~\cite{ref27} implements an enhanced Transformer model for modulation recognition and employs design space exploration (DSE) to reduce computational load and resource usage. However, the inherent model complexity introduces significant latency, constraining its applicability to real-time communication systems.

In this work, we propose two NN models, DFT-Net and Demodulation-Net (hereafter referred to as Demod-Net), designed to replace the conventional IDFT/DFT and demodulation blocks in an OFDM transceiver. The proposed framework mitigates error propagation inherent in cascaded signal-processing pipelines and minimizes the E2E bit error rate (BER) through jointly trainable models. Furthermore, several inference-oriented engineering optimizations are incorporated to enhance hardware deployability. The networks are implemented on a customized DFT–Demodulation Net Accelerator (DDNA) architecture, and experimental evaluations demonstrate that their performance satisfies the operational requirements of practical OFDM systems. The main contributions of this paper are summarized as follows:
\begin{enumerate}[label=(\arabic*)]
\item 
Unlike conventional cascaded DFT and demodulation pipelines, the proposed NNs are jointly optimized in an E2E manner, thereby mitigating inter-module error amplification while producing regular tensor shapes and fusion-friendly operator granularity suitable for fixed-point implementation. Distinct from prior DL-based PHY designs, our work incorporates a partial signal processing module of the transmitter, specifically the IDFT, into the training process, where the corresponding NN model mirrors the DFT-Net architecture. To ensure interoperability with conventional systems, we explicitly preserve an operator-equivalent path, enabling hybrid deployment with legacy modules.
\item 
The DDNA leverages the inherent parallelism and pipelining capabilities of FPGAs to enhance computation efficiency. By employing block matrix multiplication, data merging, and a fully pipelined execution schedule, the architecture achieves improved array utilization. A unified AXI interconnect is adopted to ensure compatibility with Zynq SoC platforms. Compared with generic DPUs and existing FFT IP cores, DDNA exhibits superior throughput and greater architectural flexibility under comparable latency constraints.
\item 
A complete evaluation framework encompassing algorithmic performance, model complexity, and on-hardware verification is established. Comparative analyses indicate that the proposed NN models, when deployed on the DDNA, achieve approximately 1.5 dB improvement in BER over traditional designs, with up to 66\% reduction in overall latency. The results fulfill the performance requirements of OFDM-based communication scenarios.
\end{enumerate}

The remainder of this paper is organized as follows. The Section \ref{section:System Model} presents the system model and discusses the network architecture and several optimizations. The Section \ref{section:Architecture} describes the proposed DDNA architecture. The Section \ref{section:Verification} validates and evaluates the performance of different hardware solutions. The Section \ref{section:Conclusion} concludes the paper and outlines directions for future work.

\section{System Model}\label{section:System Model}
This section first outlines a conventional OFDM system employing scattered pilots and describes the standard implementation of the DFT block at the receiver, which serves as the baseline for subsequent optimization. We then introduce two neural network–based receiver modules designed to address dynamic reconfiguration and error accumulation issues. The first, termed DFT-Net, supports flexible adjustment of the DFT size, while the second, Demod-Net, enables adaptive switching among different modulation formats. The two networks are jointly trained to minimize signal reconstruction loss, thereby improving demodulation accuracy. The external interfaces of these networks are designed to match those of conventional signal-processing blocks, ensuring compatibility for hybrid deployment. Finally, algorithmic-level optimizations are introduced to reduce computational complexity and inference latency, guided by both model architecture and hardware deployment considerations.
\subsection{OFDM System}
\begin{figure}[!t]
\centering
\includegraphics[width=3.5in]{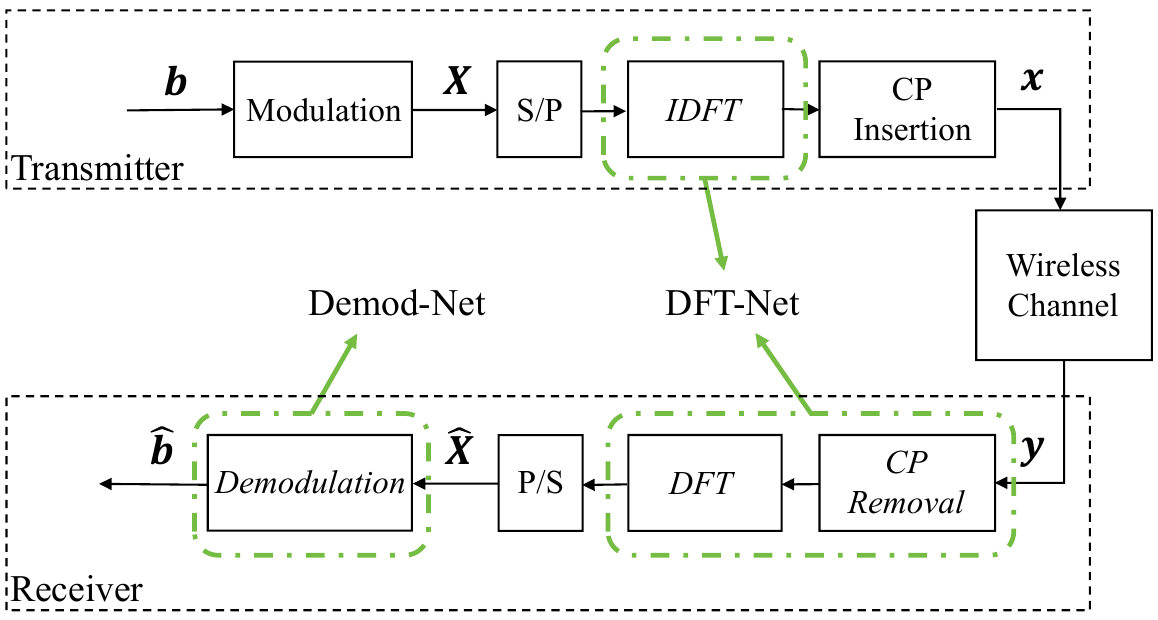}
\caption{Block diagram of a conventional OFDM system.}
\label{fig_1}
\end{figure}

The PHY architecture of an OFDM system is depicted in Fig.~\ref{fig_1}. At the transmitter, a bitstream $\boldsymbol{b}$, after channel coding, which is omitted here for clarity, is mapped to complex constellation symbols in the in-phase/quadrature (IQ) plane. Pilots and guard subcarriers are inserted, and serial-to-parallel conversion produces the frequency-domain OFDM symbol vector $\boldsymbol{X}$. An $N$-point inverse DFT (IDFT) transforms $\boldsymbol{X}$ into a time-domain waveform $\boldsymbol{x}$, to which a cyclic prefix (CP) is appended. Parallel-to-serial conversion then yields the complete OFDM time-domain signal. The CP is generated by copying the last $N_{\text{cp}}$ samples of the symbol to its beginning. The baseband signal $\boldsymbol{x}$ is upconverted to radio frequency (RF) and transmitted over the wireless channel.
\par
At the receiver, the RF waveform is downconverted to complex baseband samples and synchronized in carrier frequency and phase, recovering the time-domain OFDM signal $\boldsymbol{y}$. The CP is removed, and an $N$-point DFT is performed to obtain the frequency-domain vector $\boldsymbol{Y}$. The equalized output $\hat{\boldsymbol{X}}$ is obtained through channel estimation and compensation, which is then passed to the soft demodulator to produce probabilistic bit estimates $\tilde{\boldsymbol{b}}$. A final hard decision yields the binary bitstream $\hat{\boldsymbol{b}}$ for subsequent channel decoding.

An OFDM frame consists of $F$ symbols, each containing $N$ subcarriers. Within each symbol, $P$ subcarriers are assigned to pilots, $G$ subcarriers are reserved as guard bands (including the DC null and band-edge guards), and the remaining $D$ subcarriers carry data. The total symbol length in the time domain is $S = N + N_{\text{cp}}$.

In OFDM systems, the DFT and IDFT are the fundamental transformations between the frequency and time domains. For a sequence of $N$ complex frequency-domain samples $\{X[k]\}^{N-1}_{k=0}$, the IDFT is expressed as
\begin{equation}
x[n]= \frac{1}{N} \sum_{k=0}^{N-1}{X[k]e^{j2\pi kn/N}}, \quad n=0,...,N-1. \label{1}
\end{equation}
Conversely, the DFT (time-to-frequency mapping) at the receiver is given by
\begin{equation}
X[k]=  \sum_{n=0}^{N-1}{x[n]e^{-j2\pi kn/N}}, \quad k=0,...,N-1. \label{2}
\end{equation}

Direct matrix-based computation of these operations requires $\mathcal{O}(N^2)$ complexity, whereas FFT/IFFT algorithms, such as radix-2 or radix-4 schemes~\cite{ref4}, reduce the complexity to $\mathcal{O}(N \log N)$. The FFT achieves this efficiency by recursively decomposing the $N$-point DFT into smaller sub-DFTs, exploiting symmetry and periodicity in the twiddle factors, and reusing intermediate computations. For instance, in the radix-2 algorithm, the $N$-point sequence is divided into even- and odd-indexed subsequences, each processed by an $N/2$-point DFT. The results are then combined via a set of butterfly operations, iteratively applied until only 2-point transforms remain. This divide-and-conquer structure, coupled with intermediate result reuse, underpins both the computational efficiency and hardware friendliness of FFT architectures.

The above formulation establishes the conventional OFDM signal-processing pipeline and its underlying DFT/IDFT operations. Despite the efficiency of FFT-based implementations, they remain explicitly defined computations with fixed transform sizes and static architectures. Such rigidity imposes storage and resource constraints due to the precomputation of twiddle factors and data-reordering buffers, and results in underutilization when dynamic subcarrier configurations are required. Moreover, traditional DFT/IDFT modules cannot be co-optimized with neighboring processing blocks, leading to cumulative error propagation through the receiver chain.

To address these limitations, Section~\ref{subsec:network} introduces a learning-based operator replacement. Specifically, trainable neural networks replace selected signal-processing stages, allowing E2E co-optimization across the communication chain. The DFT-Net substitutes the conventional “CP removal and DFT” stage at the receiver, directly mapping time-domain IQ samples to frequency-domain representations. Symmetrically, it replaces the transmitter IDFT for frequency-to-time conversion. In parallel, the Demod-Net replaces the traditional soft demodulator, transforming frequency-domain symbols into soft-bit estimates. Joint training of these two networks enables unified parameter optimization across sensing and demodulation tasks, thereby improving overall signal reconstruction accuracy and robustness.

\subsection{Network Structure}\label{subsec:network}
\begin{figure}[t]  
\centering
\includegraphics[width=3.5in]{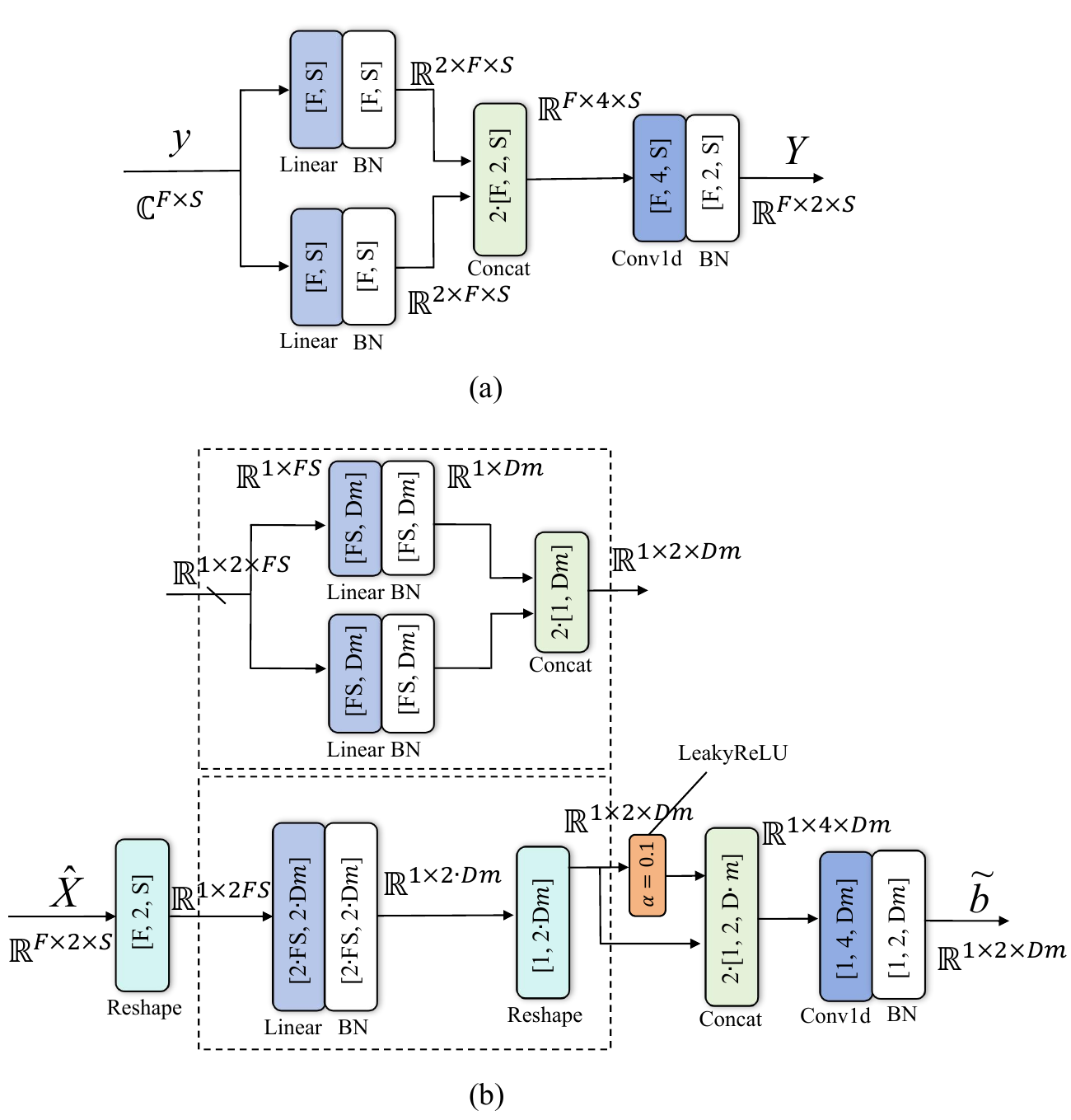}
\caption{Network architecture. (a) DFT-Net learns the bidirectional mapping between the time-domain symbol $\boldsymbol{y}$ and the frequency-domain symbol $\boldsymbol{Y}$. (b) Demod-Net provides two structural variants and performs soft demodulation from $\hat{\boldsymbol{X}}$ to soft bits $\tilde{\boldsymbol{b}}$ within one OFDM frame. Each bit is represented by two real-valued scores. The notation [$\cdot$] denotes the input dimensionality, and $m$ indicates the modulation order, e.g., $m=2$ for QPSK and $m=4$ for 16-QAM.}
\label{fig_3}
\end{figure}

The overall network architecture is depicted in Fig.~\ref{fig_3}. The proposed DFT-Net serves as a functional replacement for both the DFT and IDFT operators, with separate models trained for the transmitter and receiver. The network follows the computational structure of the conventional DFT while learning the associated transformations from data. On the receiver side, the CP segment is explicitly included as part of the learnable mapping, allowing the network to jointly capture CP-related effects.

According to \eqref{2}, the receiver-side DFT aims to recover the frequency-domain symbol vector $\boldsymbol{Y}$, expressed as
\begin{equation}
    \boldsymbol{Y} = \Re{(\boldsymbol{Y})}+j\Im{(\boldsymbol{Y})} =\mathbf{F}\boldsymbol{y},\; \mathbf{F} \in \mathbb{C}^{N\times N},\; \mathbf{F}_{k,n}=e^{-j2\pi kn/N}, \label{7}
\end{equation}
while the inverse transform is given by $\boldsymbol{y} = \mathbf{F}^H \boldsymbol{Y}$.
Decomposing $\mathbf{F}$ into real-valued cosine and sine matrices yields
\begin{equation}
    \mathbf{C}_{k,n}= \cos{\frac{2\pi kn}{N}},\quad \mathbf{S}_{k,n}= \sin{\frac{2\pi kn}{N}}, \label{8}
\end{equation}
from which the real and imaginary components of $\boldsymbol{Y}$ can be expressed as
\begin{equation}
\begin{aligned}
    \Re{(\boldsymbol{Y})} &= \mathbf{C}\Re{(\boldsymbol{y})}+\mathbf{S}\Im{(\boldsymbol{y})}, \\
    \Im{(\boldsymbol{Y})} &= 
    -\mathbf{S}\Re{(\boldsymbol{y})}+\mathbf{C}\Im{(\boldsymbol{y})}.
\end{aligned} \label{9}
\end{equation}
\par 
In DFT-Net, linear layers operate directly on the complex-valued input $\boldsymbol{y}$. Denote the equivalent weights of two Linear layers with batch normalization as $\widetilde{\mathbf{W}}_r, \widetilde{\mathbf{W}}_i \in \mathbb{R}^{S \times S}$. Four intermediate feature branches are then defined as 
\begin{equation}
\begin{aligned}
    \boldsymbol{u}_1 &= \widetilde{\mathbf{W}}_r\;\Re{(\boldsymbol{y})}, \quad 
    \boldsymbol{u}_2 = \widetilde{\mathbf{W}}_r\;\Im{(\boldsymbol{y})}, \\
    \boldsymbol{u}_3 &= \widetilde{\mathbf{W}}_i\;\Re{(\boldsymbol{y})}, \quad 
    \boldsymbol{u}_4 = \widetilde{\mathbf{W}}_i\;\Im{(\boldsymbol{y})},
\end{aligned} \label{10}
\end{equation}
which are concatenated along the channel dimension to form $[\boldsymbol{u}_1,\boldsymbol{u}_2,\boldsymbol{u}_3,\boldsymbol{u}_4] \in \mathbb{R}^{4\times S}$. A one-dimensional convolutional layer (Conv1D) with a $2\times4$ mixing kernel is subsequently applied to each sample. Let the mixing weights of the two output channels be $\boldsymbol{m}^{T}_r,\boldsymbol{m}^T_i \in \mathbb{R}^{1\times 4}$, the final outputs are given by
\begin{equation}
\begin{aligned}
    \Re{(\widehat{\boldsymbol{Y}})} &= m_{r,1}\boldsymbol{u}_1 + m_{r,2}\boldsymbol{u}_2 + m_{r,3}\boldsymbol{u}_3 + m_{r,4}\boldsymbol{u}_4, \\
    \Im{(\widehat{\boldsymbol{Y}})} &= m_{i,1}\boldsymbol{u}_1 + m_{i,2}\boldsymbol{u}_2 + m_{i,3}\boldsymbol{u}_3 + m_{i,4}\boldsymbol{u}_4.
\end{aligned} \label{11}
\end{equation}
Substituting \eqref{10} into \eqref{11} yields the equivalent transformation:
\begin{equation}
\begin{aligned}
    \Re{(\widehat{\boldsymbol{Y}})} =  &(m_{r,1}\widetilde{\mathbf{W}}_r+m_{r,3}\widetilde{\mathbf{W}}_i)\Re{(\boldsymbol{y})} \;+ \\&(m_{r,2}\widetilde{\mathbf{W}}_r+m_{r,4}\widetilde{\mathbf{W}}_i)\Im{(\boldsymbol{y})},\\
    \Im{(\widehat{\boldsymbol{Y}})} = &(m_{i,1}\widetilde{\mathbf{W}}_r+m_{i,3}\widetilde{\mathbf{W}}_i)\Re{(\boldsymbol{y})} \;+ \\&(m_{i,2}\widetilde{\mathbf{W}}_r+m_{i,4}\widetilde{\mathbf{W}}_i)\Im{(\boldsymbol{y}).}
\end{aligned} \label{12}
\end{equation}

The similarity between \eqref{9} and \eqref{12} is evident. Setting $S = N$, $\widetilde{\mathbf{W}}_r = \mathbf{C}$, $\widetilde{\mathbf{W}}_i = \mathbf{S}$, and choosing the convolutional weights as $\boldsymbol{m}_r = [1, 0, 0, 1]$ and $\boldsymbol{m}_i = [0, 1, -1, 0]$ yields exact equivalence to the conventional DFT in \eqref{9}. Similarly, setting $\boldsymbol{m}_r = [1, 0, 0, -1]$ and $\boldsymbol{m}_i = [0, 1, 1, 0]$ results in an IDFT-equivalent operation. Thus, DFT-Net is capable of emulating a broad class of DFT/IDFT mappings. In this configuration, $\widetilde{\mathbf{W}}_r$ learns a data-adaptive cosine kernel, while $\widetilde{\mathbf{W}}_i$ learns a sine kernel. During training, these weights, together with $\boldsymbol{m}_r$ and $\boldsymbol{m}_i$, are jointly optimized, generalizing the canonical DFT basis to a data-driven representation. This design preserves the inherent coupling between real and imaginary components while allowing the model to absorb scenario-dependent amplitude and phase deviations, thereby mitigating systematic error accumulation.

The Demod-Net follows a structure analogous to DFT-Net and includes two intermediate variants, namely Demod1-Net and Demod2-Net. In Demod2-Net, the real and imaginary parts of $\hat{\boldsymbol{X}}$ are processed as separate tensors, whereas Demod1-Net concatenates them into a single composite tensor. Each branch employs Linear layers with batch normalization to extract features and reshapes the output into $[B, 2, Dm]$. The LeakyReLU-activated branch is concatenated with the original tensor to form a $[B, 4, Dm]$ feature map, which is subsequently processed by a convolutional layer to generate demodulated outputs. After reshaping to $[B, Dm, 2]$, a sigmoid activation along the final dimension produces the soft bits $\tilde{\boldsymbol{b}}$, framing the demodulation task as a binary classification problem.

Demod2-Net requires only half the number of Linear-layer parameters compared with Demod1-Net. This reduction stems from the observation that when DFT-Net approximates \eqref{9}, the learned real and imaginary weights exhibit a near-diagonal symmetry. Separating the two channels thus avoids redundant feature extraction, significantly reducing parameter count and computational overhead. However, the trade-off is that independent processing may omit complementary cross-channel information, leading to less expressive feature representations.

\begin{table}[tbp]
  \centering  
  \caption{OFDM System Configuration}  
  \renewcommand{\arraystretch}{1.5}
  \begin{tabular}{|l|l|}  
    \hline  
    FFT points              & \( N = 64 \) \\
    \hline  
    Sample rate             & \( 20 \ \text{MHz} \) \\  
    \hline  
    Subcarrier spacing      & \( 312.5 \ \text{KHz} \) \\
    \hline  
    Guard SCs per symbol    & \( 8(\text{Side}) + 2(\text{DC}) \) \\  
    \hline  
    Pilot SCs per symbol    & \( P = 8 \) \\
    \hline  
    Symbols per frame       & \( F = 8 \) \\
    \hline  
    CP length               & \( 0.25N \) \\
    \hline  
    Modulation              & \( m\text{-ary QAM} \ (m=1,2,4),  \text{Gray code} \) \\
    \hline  
  \end{tabular}
  \label{tab:ofdm}  
\end{table}

\subsection{OFDM Samples and Training}\label{subsec:st}
We will evaluate the following three models: (1) Both the transmitter and receiver use traditional signal processing modules, which called conventional transceiver. (2) The transmitter still uses traditional signal processing modules, while the DFT and demodulation modules in the conventional receiver are replaced with DFT-Net and Demod-Net, called DL-receiver. (3) The traditional IDFT module in the transmitter is replaced with DFT-Net, and the receiver model is the same as in (2), called DL-transceiver. Finally, we compare the hard-decision bit outputs of the three models. 
Channel estimation and equalization are not considered in this study and will be incorporated into the receiver model in future work. In Section \ref{subsec:Analysis}, the performance of the DL-Receiver in an AWGN channel is presented for $m\text{-ary}$ QAM($m\leq 4$) modulations.
\par 
The simulated OFDM system emulates a simplified IEEE~802.11 downlink frame structure~\cite{ref28}, with system parameters summarized in Table~\ref{tab:ofdm}. The DFT size is set to $N=64$, and a total of 10 subcarriers are reserved for DC and edge guard bands. With a sampling rate of 20~MHz, the resulting subcarrier spacing is 312.5~kHz. Each frame comprises eight OFDM symbols. The numbers of pilot and data subcarriers are 8 and 368, respectively. A scattered-pilot pattern is employed, where pilot tones are fixed to $\sqrt{1/2}(1+j)$ to ensure unit magnitude. Constellation mapping adopts 2-, 4-, and 16-QAM with Gray labeling, and all constellation amplitudes are normalized to unity.

\par 

The proposed NN models are trained in an E2E manner to minimize the BER. The training SNR is randomly sampled from the range –10 dB to 30 dB. Random integer sequences from 0 to $m-1$ are generated as input data, and channel noise consistent with the target SNR is added to the transmitter output. The learning rate is initialized to 0.003. The batch size is set to 1024 during training and 512 during evaluation. The binary cross-entropy (BCE) loss is employed to measure the discrepancy between predicted and reference bits. Model parameters are optimized using the Adam optimizer~\cite{ref33}, ensuring stable convergence and improved training efficiency.

\subsection{Performance and Complexity Analysis} \label{subsec:Analysis}
In this section, we evaluate the proposed NN models under floating-point (FP) arithmetic and compare them with conventional baselines. The BER is adopted as the metric to assess E2E performance. Specifically, CP removal and DFT are replaced by DFT-Net, and the demodulation stage is replaced by Demod-Net, which together form the DL-Receiver. 

\begin{figure}[tbp]
\centering
\includegraphics[width=3.2in]{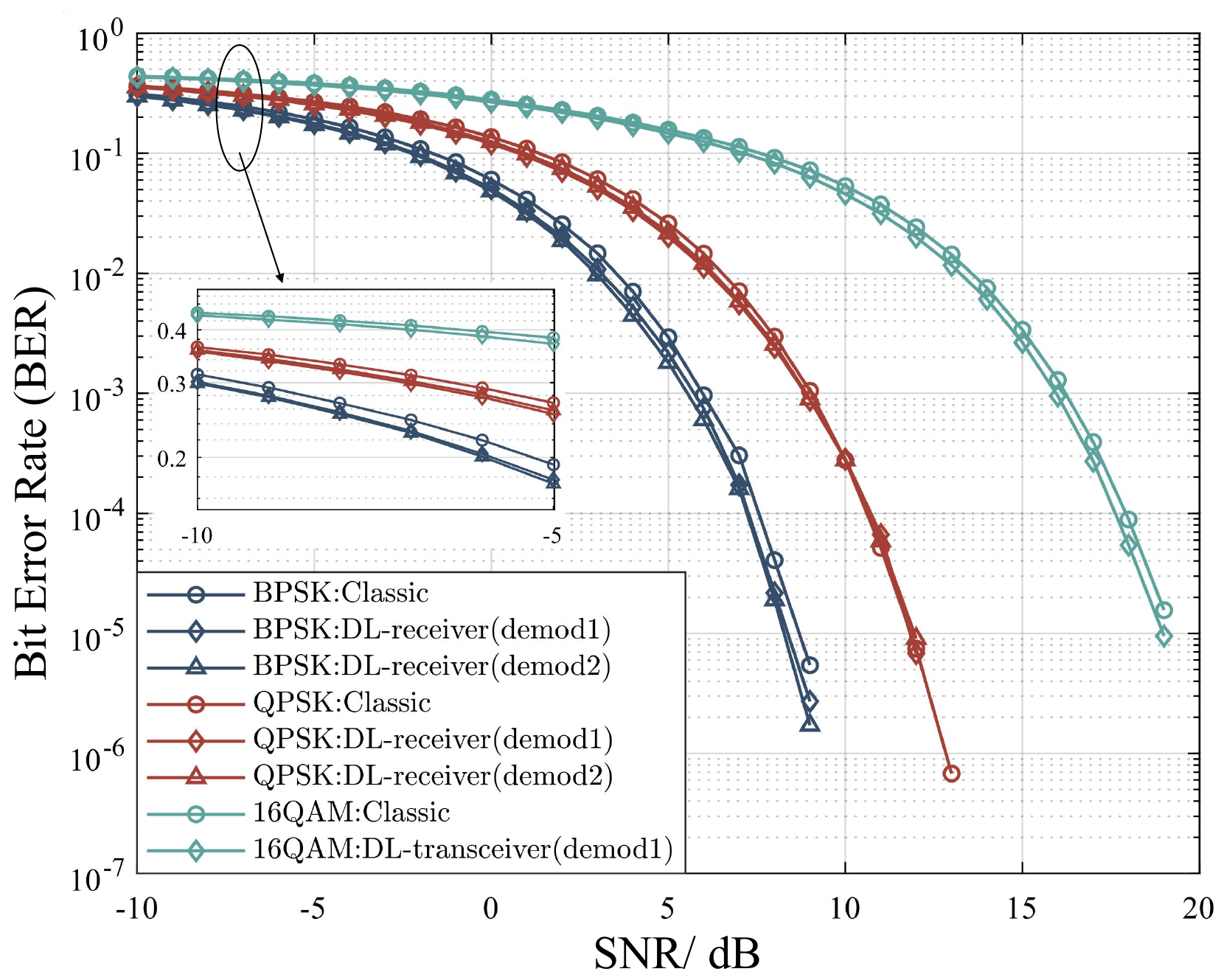}
\caption{BER performance comparison between the conventional receiver and the proposed DL-Receiver in an AWGN channel.}
\label{fig_4}
\end{figure}

Fig. \ref{fig_4} presents the BER performance in an AWGN channel for BPSK, QPSK, and 16-QAM over an SNR range from -10 dB to 20 dB, comparing the DL-Receiver with a conventional OFDM receiver. For QPSK, the overall performance is comparable, while the DL-Receiver shows gains in the 0 dB to 10 dB interval. For both BPSK and 16-QAM, the DL-Receiver achieves consistent improvements across the range. But for 16-QAM, the DL-Receiver with Demod2-Net is difficult to train to a performance level that surpasses the conventional receiver. It is worth noting that as SNR increases, BER decreases monotonically, and when it falls below $10^{-5}$ the difference becomes negligible.

\par 

\begin{figure}[tbp]
\centering
\includegraphics[width=3.2in]{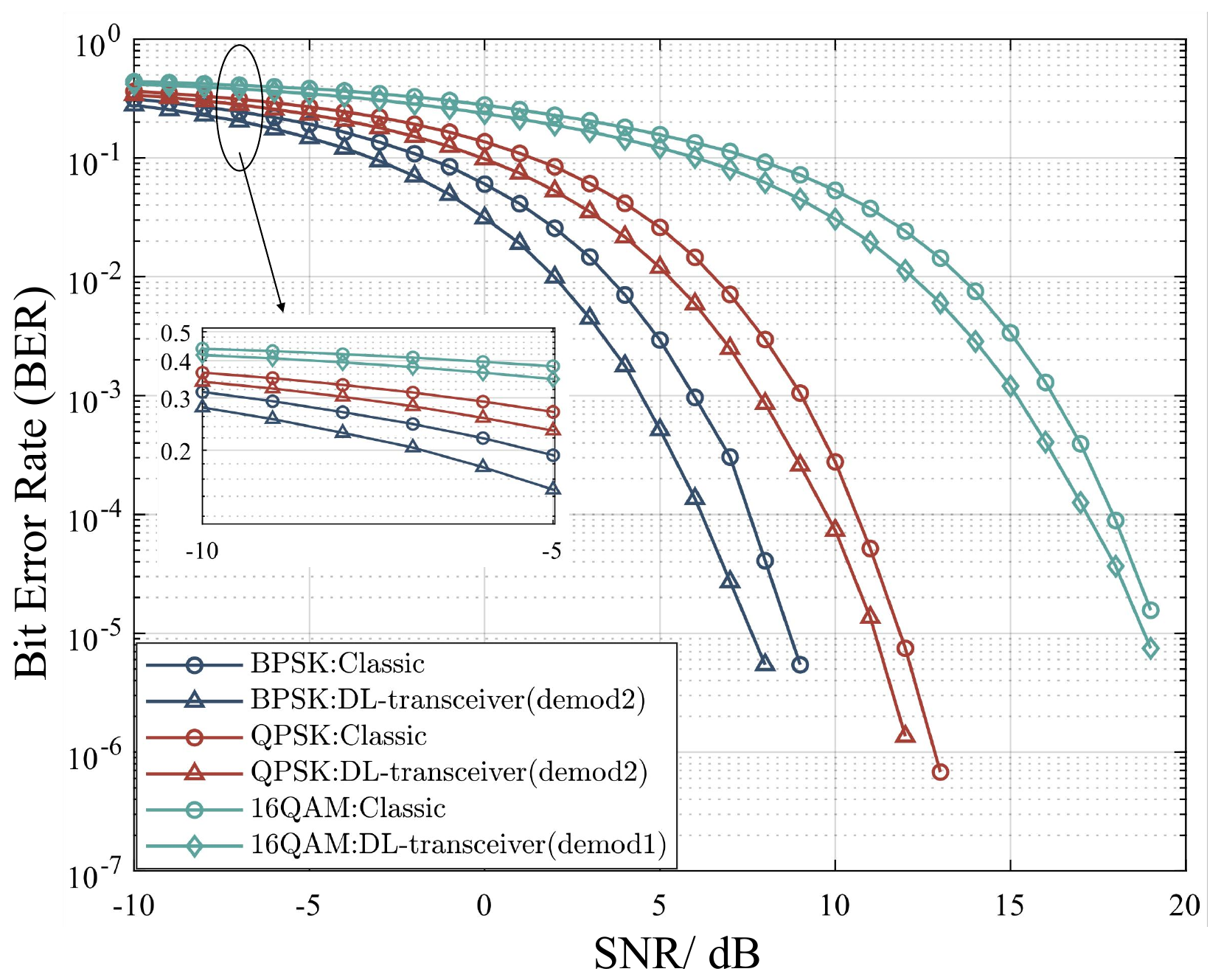}
\caption{BER performance comparison between the conventional transceiver and the proposed DL-Transceiver in an AWGN channel.}
\label{fig_12}
\end{figure}
\par 

The performance of the DL-Transceiver is shown in Fig.~\ref{fig_12}, which compares it against the conventional transceiver under identical conditions. Incorporating the DL-based IDFT at the transmitter side yields notable gains, approximately 2~dB improvement at low SNRs and around 0.9~dB at higher SNRs. These results highlight the benefits of joint end-to-end training of the transmitter and receiver modules. In practical wireless communication systems, when the modulation scheme changes, both ends of the transceiver must update the corresponding NN parameters; however, such parameter updates can be executed rapidly due to the compact model size and efficient adaptation process.

\begin{table}[t]
    \centering
    \renewcommand{\arraystretch}{1.5}
    \caption{Complexity Comparison for Models}
    \resizebox{\columnwidth}{!}{%
    \begin{tabular}{|c|c|c|c|c|c|c|}
    \hline
    \textbf{Model}         & \multicolumn{3}{c|}{DFT-Net + Demod1-Net} & \multicolumn{3}{c|}{DFT-Net + Demod2-Net} \\
    \hline
    \textbf{Modulation}             & BPSK   & QPSK   & 16QAM  & BPSK   & QPSK   & 16QAM  \\
    \hline
    \textbf{\makecell[c]{Learnable\\parameters}}   & 0.96M  & 1.89M  & 3.78M  & 0.48M  & 0.95M  & 1.89M  \\
    \hline
    \textbf{\makecell[c]{MAC\\operations}}   & 1.15M  & 2.10M  & 3.99M  & 0.68M  & 1.16M  & 2.10M  \\
    \hline
    \end{tabular}
    }
    \label{tab:Complexity}
\end{table}

\par 
To guide deployment choices, both computational complexity and link-level performance are considered. As the receiver-side NN stack is more demanding, our analysis focuses on the receiver. Table~\ref{tab:Complexity} reports the learnable parameter counts and multiply–accumulate (MAC) operations of the trainable receiver networks for different modulation orders and model variants. Parameter count primarily impacts model footprint and off-chip memory traffic, whereas MAC count reflects computational load and latency. The key distinction between the two options lies in the Linear stage of Demod-Net: Demod1-Net utilizes a single wide Linear layer, while Demod2-Net factorizes it into two narrower Linear layers, substantially reducing both parameters and MACs. Section~\ref{section:Architecture} presents a mapping architecture enabling efficient deployment of these alternatives.

\subsection{Optimization} \label{subsec:Optimization}
Section~\ref{subsec:Analysis} validated functional correctness via offline metrics under floating-point (FP) arithmetic, i.e., a prototype setting. Direct hardware realization faces two practical hurdles: (i) frequent alternation among Linear, Conv1D, and BN operators incurs extra memory transactions and scheduling overhead; and (ii) FP formats are resource-intensive and latency-prone on FPGAs. To bridge simulation and deployment, we introduce two targeted optimizations.
\subsubsection{Fusion}
Operator fusion reduces memory footprint and inference latency by collapsing adjacent operators into a single mathematically equivalent transform, thereby decreasing compute, memory accesses, and scheduling overhead. Specifically, we fold BN into the preceding Linear or Conv1D layer by absorbing BN parameters into weights and biases. The BN computation is
\begin{equation}
    \begin{aligned}
    \mu &=\frac{1}{m}\sum^m_{i=1}{x_i} & \sigma^2 &=\frac{1}{m}\sum^m_{i=1}{(x_i-\mu)^2} \\
    \hat x_i &= \frac{x_i-\mu}{\sqrt{\sigma^2+\epsilon}} & y_i &= \gamma *\hat x_i+\beta,
    \end{aligned} \label{13}
\end{equation}
where $m$ is the number of samples per output channel, $\epsilon$ is a small constant for numerical stability, and $\gamma,\beta$ are the learnable BN scale and shift. Let the preceding layer be an affine mapping $\boldsymbol{Y=WX+B}$, with BN output
\begin{equation}
    \text{BN}(\boldsymbol{Y}) =\boldsymbol{\gamma} \odot\frac{(\boldsymbol{Y})-\boldsymbol{\mu}}{\sqrt{\boldsymbol{\sigma}^2+\epsilon}} + \boldsymbol{\beta}. \\ 
 \label{14}
\end{equation}
Fusion produces an equivalent affine mapping
\begin{equation}
    \boldsymbol{Y^\prime=W^\prime X+B^\prime}, \label{15}
\end{equation}
with per-channel scale
\begin{equation}
    \boldsymbol{s}=\frac{\boldsymbol{\gamma}}{\sqrt{\boldsymbol{\sigma}^2+\epsilon}}, \label{16}
\end{equation}
so that
\begin{equation}
    \boldsymbol{W^\prime}=\text{Diag}(\boldsymbol{s})\cdot \boldsymbol{W}, \quad \boldsymbol{B^\prime}= \boldsymbol{s} \odot (\boldsymbol{B-\mu})+\boldsymbol{\beta}.  \label{17}
\end{equation}

\subsubsection{Quantization}
\begin{figure}[tbp]
\centering
\includegraphics[width=3.4in]{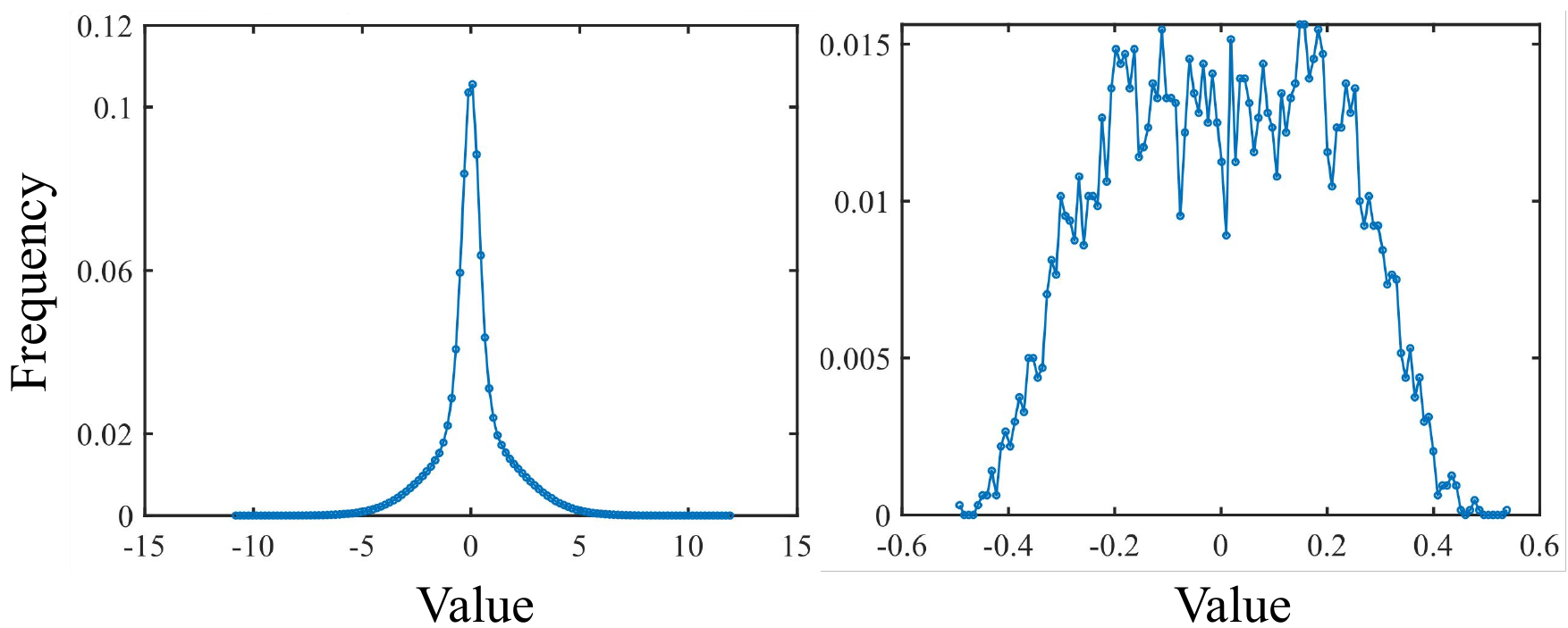}
\caption{Frequency distribution of activations and weights. The left figure represents the activations, the other represents the weights. }
\label{fig_5}
\end{figure}

FP arithmetic is costly in FPGA fabric and introduces additional latency. Prior work~\cite{ref29} indicates that quantizing weights and activations to INT8 usually incurs only modest accuracy loss. Quantization employs a scale and zero-point to approximate FP values. Parameters can be derived using min–max or histogram-based calibration; the latter is preferable for concentrated or heavy-tailed distributions. Fig.~\ref{fig_5} shows representative activation/weight histograms with strong central mass and extended tails, for which min–max calibration would underutilize the central precision. Similar patterns appear across layers.

\par 
We quantize weights and activations to INT8 using histogram-based calibration, shrinking model size to $\approx!1/4$ and reducing storage bandwidth. Because tensors typically require different scale/zero-point pairs, we adopt an execution scheme that unifies integer arithmetic and dequantizes only at interfaces. Let $S,Z$ be scale and zero-point, and $x,q$ be FP and quantized values:
\begin{equation}
    x=S_{out}(q_{out} - Z_{out})=\sum S_{in}(q_{in}-Z_{in})*S_{w}(q_{w}-Z_{w})+bias. \label{18}
\end{equation}
\par 
Guided by Fig.~\ref{fig_5}, we use symmetric quantization ($Z=0$). All operators use INT8 paths with high-precision accumulation and requantization:
\begin{equation}
    q_{out}=\frac{S_{in}*S_{w}}{S_{out}} \sum q_{in}q_{w}+\frac{S_{in}*S_{w}}{S_{out}} \cdot \frac{bias}{S_{in}*S_{w}}.  \label{19}
\end{equation}
The bias is quantized to INT16 with scale $S_b=S_{in}S_w$. In the final computation, the requantization factor $S_{in}S_w/S_{out}$ is constant across invocations but is FP, so it must be quantized to enable an integer-only pipeline.
\begin{equation}
    \frac{S_{in}*S_{w}}{S_{out}} \simeq m\cdot2^{-n},  \label{20}
\end{equation}
where each Linear/Conv1D layer has its own $(m,n)$ pair configured in registers (see Section~\ref{section:Architecture}). These optimizations condense the network into an inference-ready graph with a lower footprint and bandwidth while preserving accuracy and offering a unified numeric interface for hardware.

\section{Architecture Mapping on SoC}\label{section:Architecture}
We now map the optimized DFT-Net and Demod-Net to a heterogeneous SoC. Top-level modules expose AXI interfaces for data and instruction paths to streamline system integration. Prototyping and validation are performed on a Xilinx Zynq platform comprising a processing system (PS) and programmable logic (PL) connected via AXI. The PS orchestrates control and data movement; the PL executes parallel compute, satisfying real-time and configurability requirements.

\begin{figure}[t]
\centering
\includegraphics[width=3.4in]{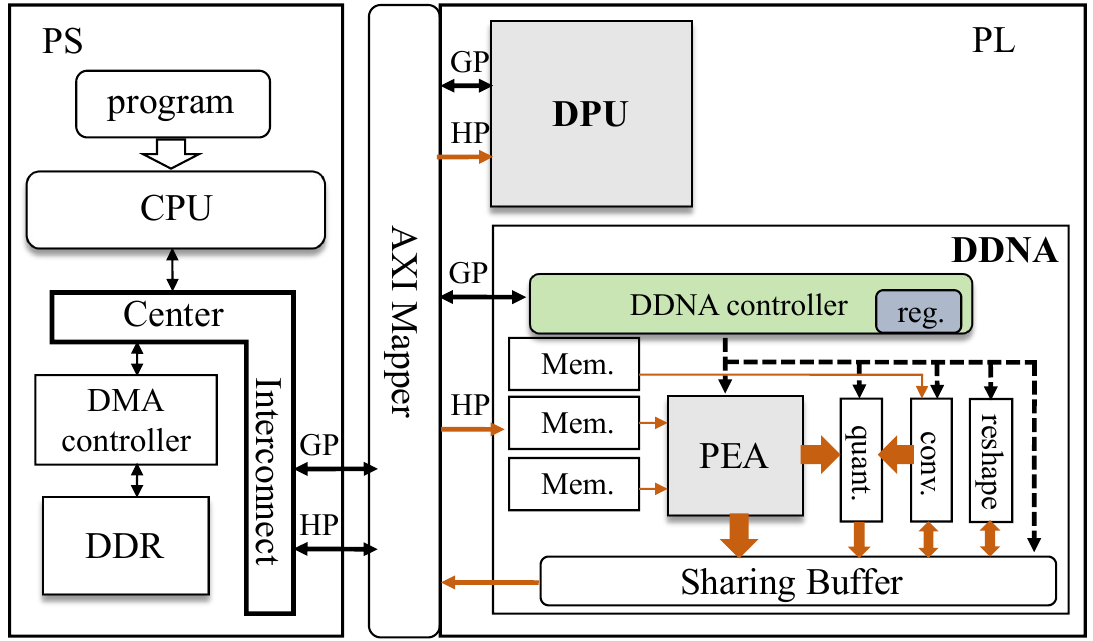}
\caption{SoC-level overview of DPU (overlapping design) and DDNA (streaming design).}
\label{fig_6}
\end{figure}

\subsection{Mapping Patterns}
FPGA accelerators generally follow overlapping or streaming designs. The overlapping style uses a general compute unit for diverse operators (e.g., FC and Conv) with broad adaptability but lower throughput; the streaming style employs specialized units to maximize efficiency.

We implement both patterns (Fig.~\ref{fig_6}). For the overlapping approach, we adopt Xilinx Vitis AI~\cite{ref30} to deploy models from PyTorch~\cite{ref31}/TensorFlow~\cite{ref32} onto the PL via the Deep Learning Processor Unit (DPU). For streaming, we develop a custom DDNA accelerator with dedicated compute units and model-tailored optimizations to maximize resource utilization.

\begin{algorithm}[!t] 
  \caption{DDNA-Oriented Control Flow} 
  \label{alg:conv_flow} 
  \LinesNumbered 
  \SetKwInOut{Input}{Input} 
  \SetKwInOut{Output}{Output} 
  \Input{Data and weights in DDR}
  \Output{Final convolution results (stored in DDR)} 
  
  \textbf{Step 1: PS Initialization \& Data Migration}\;
  \Indp 
    1.1 Configure PL-side controller registers via PS\;
    1.2 Use DMA to transfer data and weights from DDR to PL memory\;
  \Indm 
  \textbf{Step 2: DDNA Activation \& Data Calculation}\;
  \Indp
    2.1 Enable \texttt{start\_signal} to let DDNA start computation\;
    2.2 PEA performs matrix calculation\;
    2.3 \uIf{First calculation}{ 
      2.3.1 Goto \textbf{Step 3}\;
    }
    \Else{ 
      2.3.1 Wait for completion of previous convolution (Conv1D unit pulls high \texttt{conv\_done} after calculation)\;
      2.3.2 PS polls \texttt{conv\_done} until it is high\;
      2.3.3 PS controls DMA to transfer computed results back to DDR\;
      2.3.4 PS polls \texttt{recv\_done} (pulled high after DMA transfer) until it is high\;
    }
  \Indm
  \textbf{Step 3: Signal Reset \& Loop Control}\;
  \Indp
    3.1 PS pulls low \texttt{start\_signal}, \texttt{conv\_done} and \texttt{recv\_done} signals\;
    3.2 PS polls \texttt{array\_done} until it is high (PEA pulls high the signal after calculation. And Conv1D unit will operate automaticly after PEA finished)\;
    3.3 Back to \textbf{Step 1}\;
  \Indm
\end{algorithm}

\begin{figure*}[!t]
\centering
\includegraphics[width=5in]{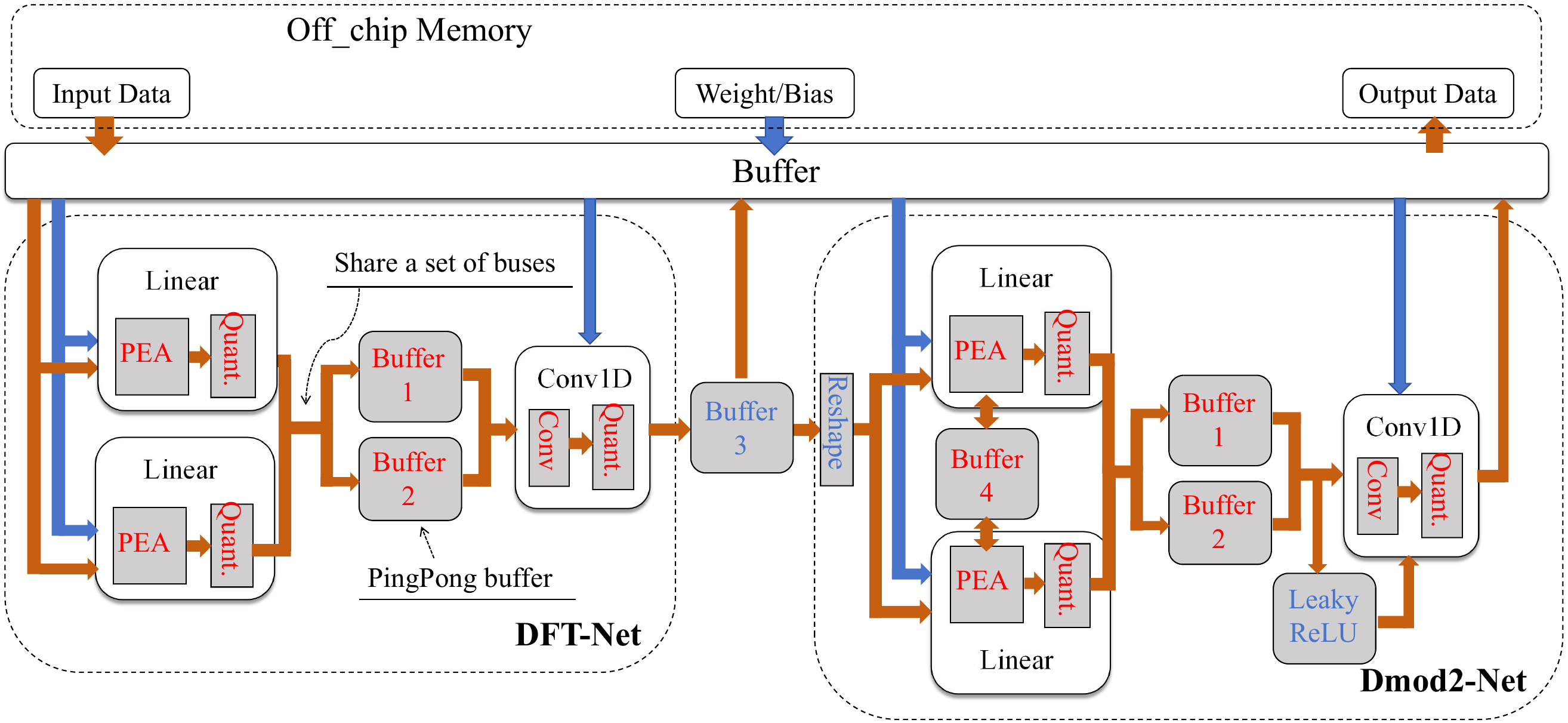}
\caption{DDNA execution trace validating functional correctness.}
\label{fig_7}
\end{figure*}

In Fig.~\ref{fig_6}, brown lines indicate dataflow and black lines instruction/control. DDNA comprises a processing element array (PEA), Conv1D unit, quantization, reshape, on-chip memory, and auxiliary blocks. The PEA implements Linear-layer MACs; Conv1D performs 1D convolutions. Quantization converts accumulations to INT8 for uniform precision and intermediate storage; reshape formats outputs on PL to reduce latency. PS/PL instructions are staged via controller registers. Data registers store per-layer requantization factors (parameter $m$ in~\eqref{20}); control registers manage I/O enables, module modes/flows, and execution status signals.

Algorithm~\ref{alg:conv_flow} summarizes PS–PL interaction from initialization and DMA staging to compute and reset. Status signals (e.g., \texttt{array\_done}, \texttt{conv\_done}) are polled for flow control.

\subsection{DDNA Architecture}
Using DFT-Net and Demod2-Net (Sec.~\ref{subsec:network}) as examples, Fig.~\ref{fig_7} illustrates mapping from algorithmic modules to DDNA hardware blocks. Receiver-side IQ samples and model weights reside in off-chip DDR to conserve on-chip resources. Brown arrows denote input-data movement; blue arrows denote weight movement. Blue blocks are dedicated units; red blocks are reusable units. The controller invokes units iteratively to complete the pipeline.

During execution, the PEA and Conv1D units fetch their (weight, bias) pairs. PEA inputs originate from on-chip memory or Buffer~3 (DFT-Net output). Linear layers execute serially on a shared PEA and write results to Buffer~1/2 configured as ping–pong buffers to avoid overwrite and boost parallelism. Conv1D reads from the complementary buffer. When DFT-Net operates as an IDFT, its output is routed to DDR. For Demod2-Net, due to heavier Linear layers, Buffer~4 stores intermediate results. Conv inputs pass through LeakyReLU; the associated dataflow is governed by controller registers (Step~1, Algorithm~\ref{alg:conv_flow}).

\subsection{Matrix Block Operations}
The NN models involve a large number of matrix multiplications, which would be resource-intensive if implemented one by one. We therefore partition matrices into blocks and decompose each large matrix multiplication into accumulations of small block multiplications. These block multiplications are mapped onto the PEA. The PEA adopts an output-stationary systolic array architecture {\cite{ref29}}, which provides efficient local data movement and high parallelism, thereby achieving excellent hardware utilization. 

For the matrix multiplication $A^{m \times s}$ and $B^{s \times n}$, to achieve low latency and high utilization on the PEA, the array dimensions should be factors of the problem size: the number of rows should divide $m$, and the number of columns should divide $n$. We set the PEA row count to $2kF$ with 
$k \in \mathbb{N}^+$, and the column count to the greatest common divisor (GCD) of columns of all weight matrices that is $16$. The GCD is chosen to improve PEA resource utilization and computational parallelism. For the Linear layer, the GCD of the input matrix row count equals 2F. Hence we introduce a parameter k to concatenate matrices, which provides a tunable tradeoff between resource usage and latency. For example, in the Linear layer of the DFT-Net, the matrix multiplication $[2F,S]\times[S,S]$ must be executed $S/16$ times sequentially on the PEA to obtain the complete result. Since the computation is partitioned along the column dimension, no intermediate buffering is required.

\begin{figure}[t]
\centering
\includegraphics[width=3.6in]{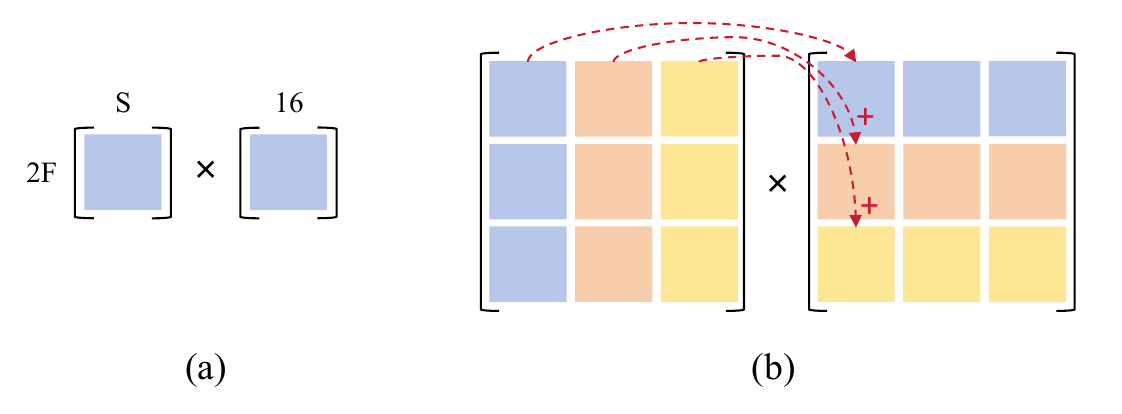}
\caption{(a) Basic block multiplication. (b) Decomposition of matrix calculation including matrix multiply and accumulate. }
\label{fig_8}
\end{figure}

Matrix block operations are illustrated in Fig. {\ref{fig_8}}, the matrix handled per PEA invocation is referred to as a basic block. A large matrix multiplication can be decomposed into multiplications and additions over multiple basic blocks, which is straightforward. To avoid extra intermediate buffering that would consume hardware resources, partial final results should be produced as early as possible and then written back to off-chip memory. For example, the two large Linear layers in Fig. {\ref{fig_8}} are decomposed into $3\times3$ basic blocks. During computation, a $1\times3$ row vector is multiplied by a $3\times1$ column vector, with elementwise products accumulated to form the output. Under this configuration, only the results for a single basic block multiplication need to be buffered. Blocks decomposing only affects Step 2.2 in Algorithm \ref{alg:conv_flow}. For the first block, the PEA stores the computed results in buffer 4 shown in Fig. \ref{fig_7} without output. For intermediate blocks, the previous partial results are retrieved from buffer 4, accumulated with the current results, and written back. For the final block, the accumulated results are output and trigger the subsequent modules.

\subsection{Data Merge} \label{subsec:merge}
\begin{figure}[t]
\centering
\includegraphics[width=3.2in]{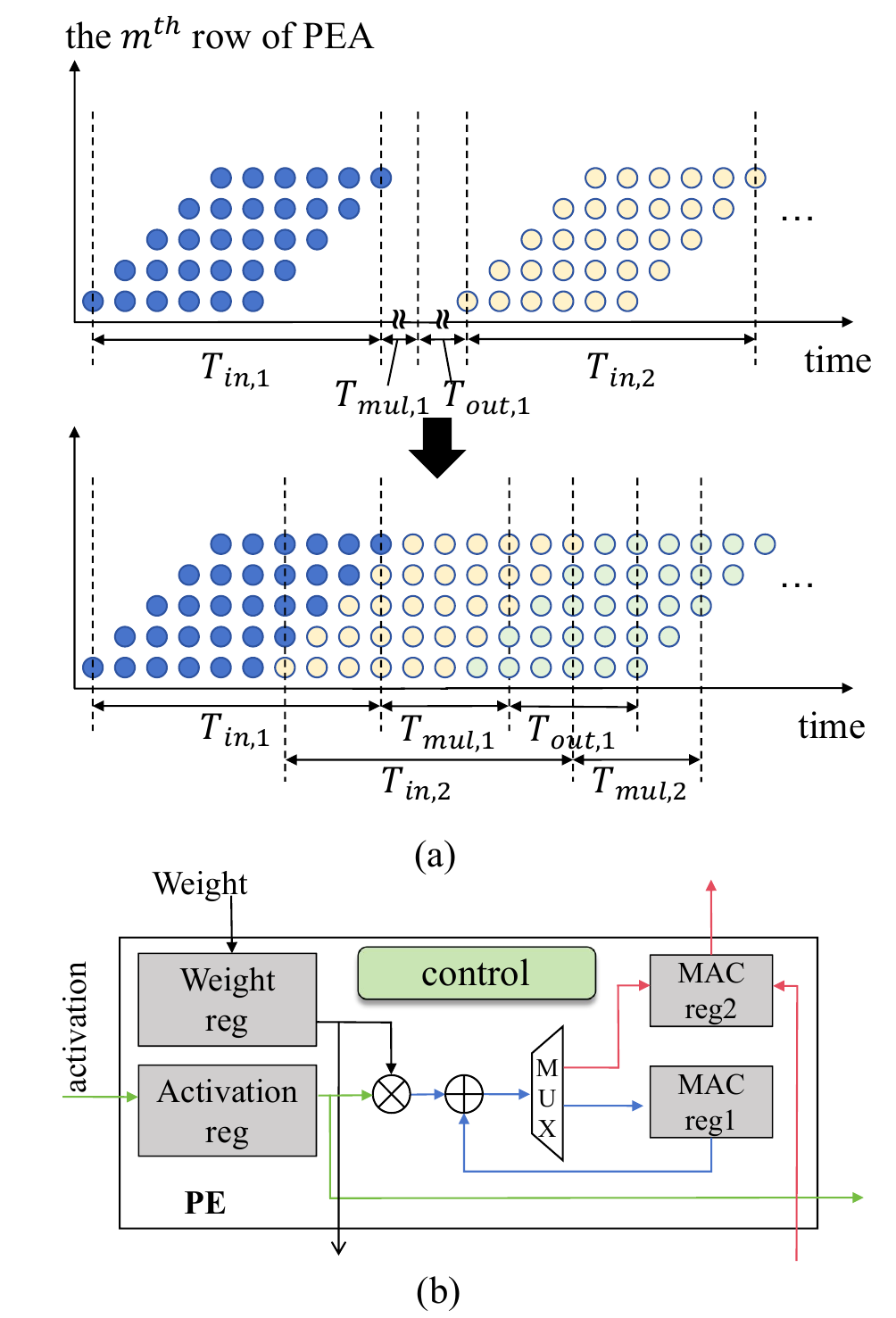}
\caption{(a) Merged input dataflow for the PEA, converting processing from serial to parallel. (b) Schematic of the modified PE internals.}
\label{fig_9}
\end{figure}

In addition, we redefine the PE of PEA to improve computational efficiency. For an output-stationary systolic array, the temporal arrival pattern of input data forms a diamond shape. After all data required by a matrix multiplication have been injected into the PEA, an additional $T_{mul}$ is needed to finish the computation. Moreover, at the end of each multiplication, the results must be read out from the internal PE registers before the next multiplication can start, which incurs a latency of $T_{out}$. To address these overheads, we propose a design with high parallelism across computation, storage, and readout, as illustrated in Fig. {\ref{fig_9}}. 

Assume that on a $5\times5$ PEA we repeatedly execute the matrix multiplication $A^{5 \times6}B^{6\times5}$. The input time is $T_{in}$, and the time from the arrival of the last input to completion is $T_{mul}$. Consider that the output readout latency $T_{out}$ is short. If data input, multiplication, and output readout are executed serially, substantial idle time appears between stages, which reduces resource utilization, as shown in the upper plot of Fig. {\ref{fig_9}}(a). We therefore merge the data paths so that while the previous matrix operation is being executed and its results are read out, inputs for the next matrix can be injected concurrently. This overlapping of computation, readout, and input is illustrated in the lower plot of Fig. {\ref{fig_9}}(a). For consecutive matrix multiplications, the latter approach can effectively reduce the overlapping computation time. To prevent data conflicts, we modify the internal structure of the PE, as shown in Fig. {\ref{fig_9}}(b). Two result registers are introduced inside the PE, and a control module manages results storage and forwarding. For example, when MAC reg1 stores the results of the current matrix operation, the final multiply–accumulate output should be written to MAC reg2, and MAC reg1 should be cleared to avoid contaminating the next matrix computation. In parallel, the data in MAC reg2 must be read out before the next set of results is produced to avoid being overwritten.

We employed a systolic array to build the PEA, and the closer the input matrix dimensions are to the PEA dimensions, the higher the computational efficiency. It should also be noted that the Linear layer in Demod-Net takes a row vector as input. When mapped to the PEA, typically only the first row of the array is active while the remaining PEs remain idle, which reduces resource utilization and increases latency. A practical approach is, for an already trained model, internal weights remain fixed across multiple inferences, which allows us to exploit the local dataflow of the systolic array for extensive data reuse. Prior to executing Demod-Net, $2kF$ input groups are buffered. Once all required data have been written, inputs are fed so that each row of the systolic array receives one group, rather than activating only the first row.

\subsection{Parallel Strategy} \label{subsec:ps}
 In DDNA, several latency components arise at different stages, including memory read/write latency, matrix calculation latency in PEA, convolution latency in the Conv1D unit, intermediate buffering latency in sharing buffer, and result write-back latency. These latencies accumulate and significantly increase the total processing time. Establishing a fully pipelined architecture is the first step toward mitigating this problem. As shown in the DDNA architecture of Fig. \ref{fig_6}, we decouple convolution from matrix operations and design a dedicated convolution unit to increase parallelism. Since matrix computations in Linear layers are more time-consuming than other operations, an effective strategy is to run operators that do not interfere with the matrix engine yet facilitate forward inference during this period. Accordingly, we partition data processing on DDNA into five classes, registers configuration of controller, input loading, matrix computation, convolution computation, and output processing, and schedule them as a time-domain pipeline.
 
 \begin{figure}[t]
\centering
\includegraphics[width=3.5in]{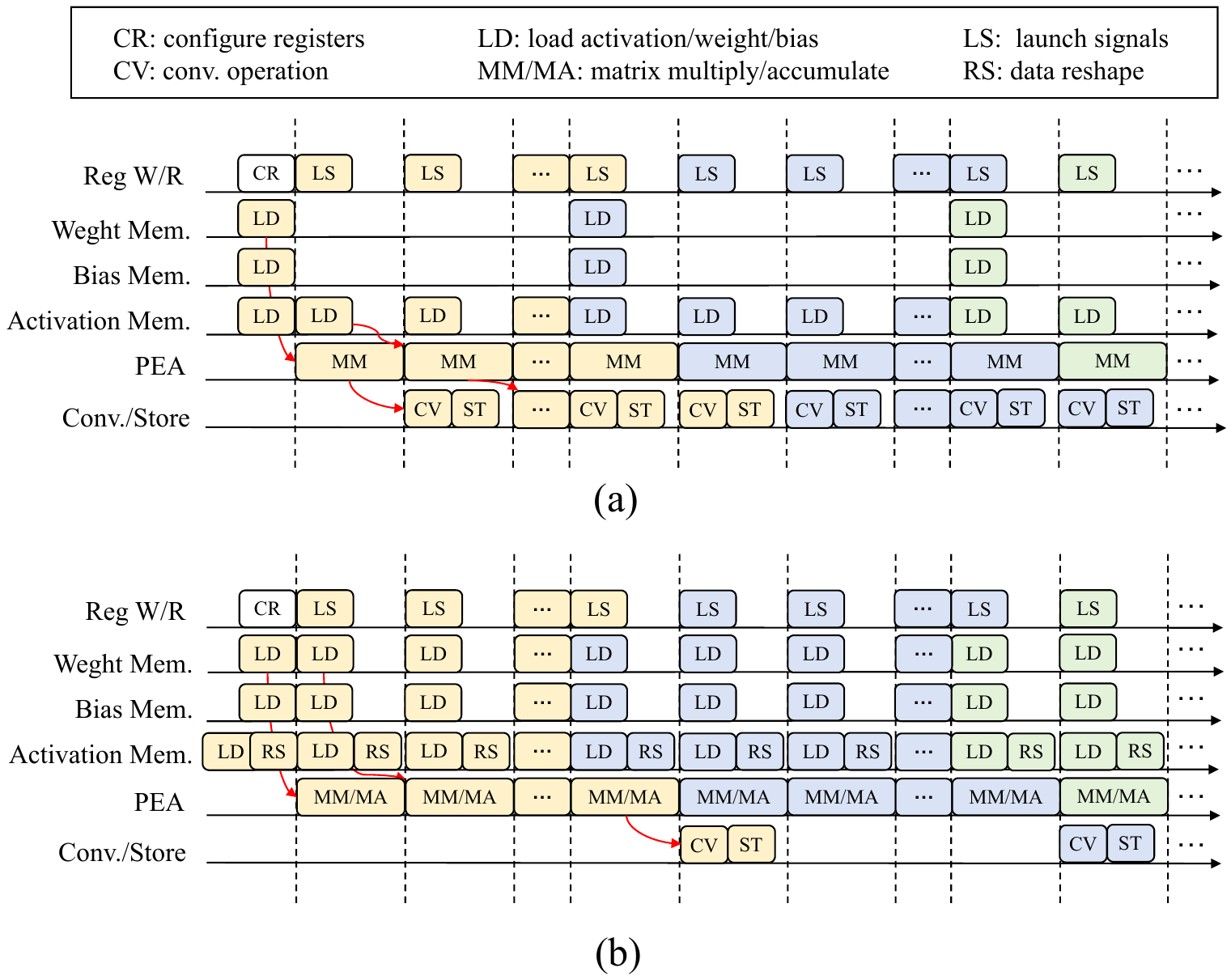}
\caption{Timing diagrams of fully pipelined operation on DDNA for (a) DFT-Net and (b) Demod-Net.}
\label{fig_10}
\end{figure}

As shown in Fig. {\ref{fig_10}}, DFT-Net and Demod-Net share similar pipeline stages on DDNA, although their procedural details differ. Fig. {\ref{fig_10}}(a) depicts the pipeline of DFT-Net. In the first stage, as described in Algorithm \ref{alg:conv_flow}, the PS initializes the registers in the DDNA controller and preloads the required data. Multiple sets of matrices containing the data of two complete Linear layers are fed in. During a single MM time, multiple consecutive matrix multiplications, data stream consistent with the lower plot of Fig. \ref{fig_9}(a), are performed to obtain the final results of the two Linear layers in the DFT-Net. If subsequent operations continue on the same Linear layer, the weights and biases do not need to be reloaded, which improves data reuse. In the second stage, the start signal is asserted and the data loaded in the first stage are processed on the PEA for matrix multiplication, after which the results are quantized to INT8. This stage is the most time-consuming. In the final stage, the matrix multiplication output is processed by the convolution unit, quantized, and then written to the buffer for downstream use. Fig. {\ref{fig_10}}(b) shows the pipeline for Demod-Net, where a reshape operation is required before data input. Unlike (a), the Linear layer is larger and cannot be completed in a single pass, so intermediate results must be buffered and multiple basic-block multiplications and additions are performed. The output is produced only after the final block accumulation. Moreover, the weights required by successive basic-block multiplications differ and cannot be reused, so each pipeline round reloads the weight data.

\section{Verification and Evaluation}\label{section:Verification}
This section focuses on hardware deployment and evaluation, covering the DL-Transceiver, the DL-Receiver, and a conventional baseline. First, the relative performance ordering among receivers is largely insensitive to whether the transmitter operates in FP or fixed/quantized precision. Second, the transmitter’s core signal-processing blocks mirror those in the receiver; duplicating transmitter deployment would therefore be redundant. Our primary objective is to verify that a DL-based hardware PHY can satisfy OFDM requirements. To this end: (i) the receiver is more complex, making it the appropriate target to demonstrate the DDNA architecture (Sec.~\ref{section:Architecture}); (ii) the DL-Transceiver employs the same IDFT (DFT-Net), which directly maps to DDNA without additional design effort; and (iii) the receiver model is common to both DL-Transceiver and DL-Receiver, so changing the transmitter requires only updating receiver NN parameters, without hardware modifications. Consequently, the transmitter is fixed in FP and implemented in software for ease of testing.

Experiments are conducted on the Xilinx ZCU106, a Zynq heterogeneous SoC with a quad-core ARM Cortex-A53 as the PS and an UltraScale+ FPGA as the PL. Available resources include 624 blocks of 18 KB RAM, 1728 DSP48E slices, 230400 six-input LUTs, and 460800 flip-flops. We also investigate the resource–latency tradeoff using the DDNA architecture. We further examine the resource–latency trade-off enabled by DDNA.

\begin{figure}[t]
    \centering
    \includegraphics[width=3.4in]{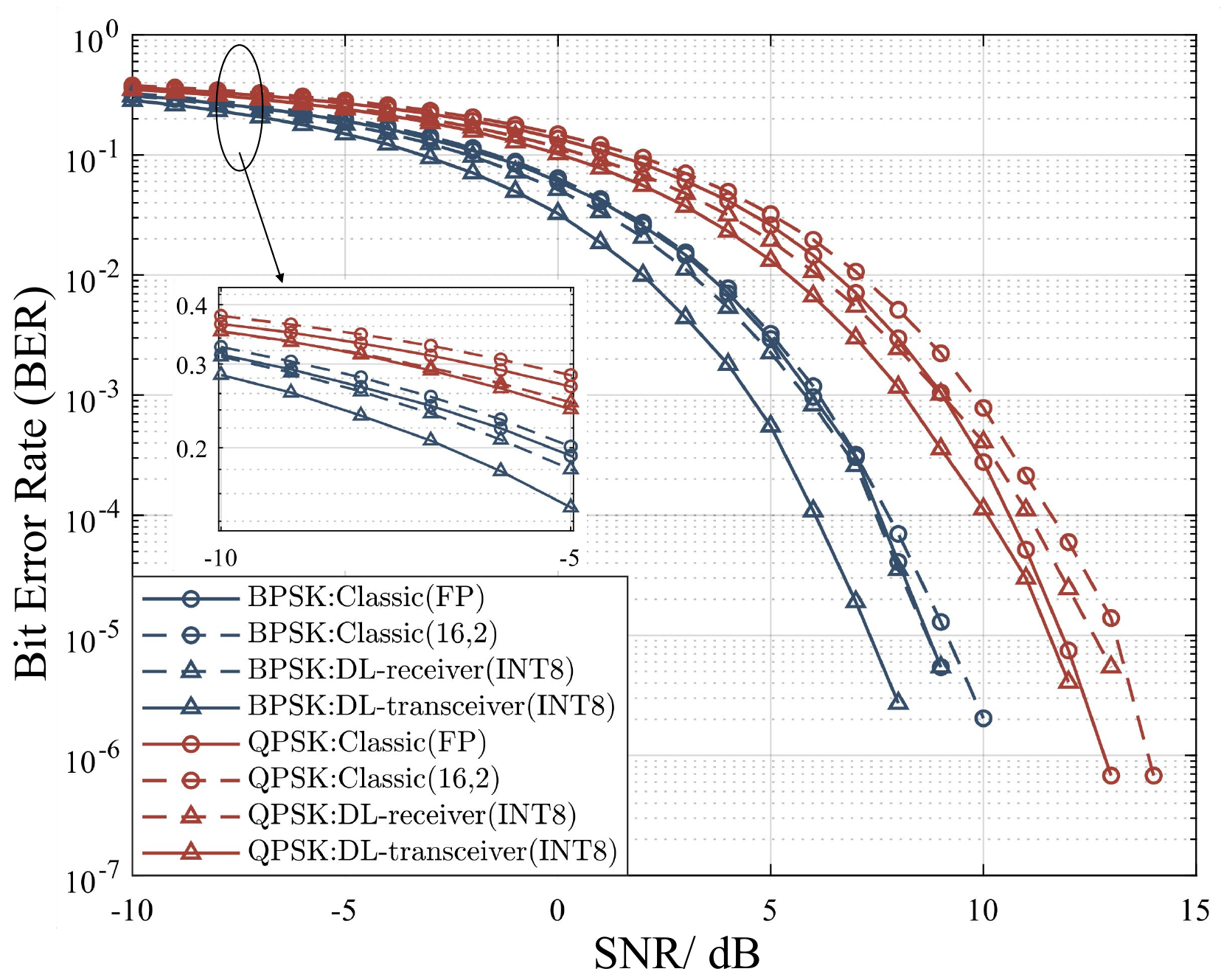}
    \caption{BER performance comparison of conventional FFT (FP and fixed point) and the INT8 NN models. Demod2-Net is used in NN models.}
    \label{fig_11}
\end{figure}

\begin{table*}[!t]
    \centering
    \renewcommand{\arraystretch}{1.5}
    \caption{The Comparison of Hardware Implementations for FFT And DFT-Net}
    \begin{tabular}{|c|c|c|c|c|c|c|c|c|c|c|c|}
    \hline
    \textbf{Platform}  & \textbf{\makecell[c]{Frequency\\(MHz)}}   & \textbf{\makecell[c]{Data Width\\(bit)}}  & \textbf{\makecell[c]{Latency\\(us))}}  & \textbf{\makecell[c]{Rate\\(MS/s)}}  
    & \textbf{\makecell[c]{Throughput\\(GOPS)}}  & \textbf{BRAM}  & \textbf{DSP}  & \textbf{FF} 
     & \textbf{LUT}  & \textbf{\makecell[c]{Power\\(W)}}  & \textbf{\makecell[c]{Energy/S\\(J/S)}} \\
    \hline
    Xilinx FFT IP  & 200   & 16   & 1.24  & 51.6   & 1.31   & 55 & 9 & 3182 & 1782 & 0.2 & $3.8\times10^{-9}$  \\
    \hline
    \cite{ref8}   & 200  & 16  & 0.40  & 160.0  & 4.08  & 0 & 24 & 1894 & 2732 & - & -  \\
    \hline
    \cite{ref36}  & 200  & 16 & 0.32 & 200.0 & 5.10 & 8 & 0 & 1941 & 10637 & 0.45 & $2.2\times10^{-9}$  \\
    \hline
    \cite{ref10}  & 200  & 16 & 0.14 & 457.1 & 11.65 & 148 & 192 & 34441 & 13897 & - & -  \\
    \hline
    DDNA($16\times16$) & 250 & 8 & 0.46 & 173.9 & 115.21 & 45 & 312 & 18772 & 13192 & 0.77 & $4.4\times10^{-9}$ \\
    \hline
    DDNA($32\times16$) & 250 & 8 & 0.25 & 320.0 & 208.46 & 45 & 568 & 33102 & 23654 & 1.86 & $5.8\times10^{-9}$ \\
    \hline
    \end{tabular}
    \label{tab:fft}
\end{table*}

\subsection{Comparison of Performance on DFT}
For the hardware deployment of the conventional FFT, we use the Xilinx FFT IP core {\cite{ref11}}. The input format is fixed point with $(16,2)$, where $W=16$ is the total bit width and $I=2$ is the number of integer bits. To validate quantized NN behavior at scale, we emulate INT8 inference in software with quantize/dequantize applied only at the module under test; all other receiver blocks remain FP, ensuring bit-exact parity with hardware.

Using the OFDM frame in Table~\ref{tab:ofdm}, Fig.~\ref{fig_11} reports BER under AWGN for BPSK and QPSK (Demod2-Net as demodulator). In the conventional receiver, fixed-point FFT exhibits a minor loss versus FP under BPSK, and an increasing gap for QPSK at higher SNR. For the DL-Receiver, the INT8 model matches or exceeds FP FFT at low SNR; at high SNR, the BER floor is dominated by quantization and finite test length (e.g., $\le 10^{-5}$). The DL-Transceiver outperforms the fixed-point conventional baseline for both BPSK and QPSK and surpasses the FP baseline. Similar trends hold for other modulations.

Table~\ref{tab:fft} summarizes 64-point DFT hardware, including latency, rate, throughput, and utilization (radix-4 FFT for prior art). The processed sample rate (subcarriers/s) is
\begin{equation}
    Rate=\frac{1}{Lat.} \times N(S). \label{21}
\end{equation}
Here, $N=64$ for FFT and $S=80$ for NN models. Throughput counts one MAC as two operations. We map DFT-Net onto DDNA with two PEA sizes ($16\times 16$ and $32\times 16$) and normalize latency by feeding identical input volumes. Conventional FFTs offer tunable trade-offs among complexity, latency, resources, and flexibility. The Xilinx IP achieves $1.24~\mu$s at 200~MHz with very low resources but fixed size. Pipeline parallelism~\cite{ref8} reduces latency; dynamic reconfigurability~\cite{ref10,ref36} supports 16–1024 points; \cite{ref36} further replaces precise multipliers with shift–add to reduce DSP usage, while \cite{ref10} attains the lowest latency at higher resource cost.

Unlike FFT variants, which compute \eqref{2} exactly and therefore cannot improve BER, DFT-Net on DDNA reaches up to 320.0~MS/s and, due to larger operation counts, attains hundreds of GOPS (over an order of magnitude above FFT IPs), underscoring DDNA’s suitability for NN workloads. DDNA($32\times16$) delivers nearly twice the throughput of DDNA($16\times16$). Power increases with array size, and energy per sample (E/S) rises accordingly; DDNA($32\times16$) exhibits $\approx1.5\times$ the E/S of the Xilinx FFT, reflecting the higher compute intensity of NN workloads. By contrast, \cite{ref36} lowers E/S via sparse twiddle-factor representations. In short, while NN models demand more compute than FFT, INT8 quantization curbs complexity and bandwidth while still yielding BER gains.

\begin{table*}[!t]
    \centering
    \renewcommand{\arraystretch}{1.5}
    \caption{The Comparison of Hardware Implementation of conventional and DL-based Receivers}
    \begin{tabular}{|c|c|c|c|c|c|c|c|c|c|c|c|}
    \hline
    \textbf{Platform}  & \textbf{\makecell[c]{Frequency\\(MHz)}} & \textbf{\makecell[c]{Data Width\\(bit)}} & \textbf{\makecell[c]{With\\Model}}  & \textbf{\makecell[c]{Latency\\(us))}}  & \textbf{\makecell[c]{Throughput\\(GOPS)}}  & \textbf{BRAM}  & \textbf{DSP}  & \textbf{FF} 
     & \textbf{LUT}  & \textbf{\makecell[c]{Power\\(W)}}  & \textbf{\makecell[c]{GOPS/W}} \\
    \hline
     Conventional & 20 & 16 & - & 32.00 & - & 10.5 & 12 & 5406 & 4162 & 0.59 & - \\
    \hline
    DPU   & 200 & 8 & Demod1 & $3.5\times10^{4}$  & 0.13 & 255 & 710 & 98118 & 50499 & 1.99 & 0.07\\
    \hline
    \multirow{2}{*}{DDNA($16\times16$)} & \multirow{2}{*}{250} & \multirow{2}{*}{8} & Demod1 & 39.44  & 114.82 & \multirow{2}{*}{45}  & \multirow{2}{*}{312} & \multirow{2}{*}{18772} & \multirow{2}{*}{13192} & \multirow{2}{*}{0.77} & 149.11 \\
    \cline{4-6} \cline{12-12}
            &   &  & Demod2 & 21.68  & 114.42 & & & & & & 148.59 \\
    \hline
    \multirow{2}{*}{DDNA($32\times16$)} & \multirow{2}{*}{250} & \multirow{2}{*}{8} & Demod1 & 19.73  & 229.53 & \multirow{2}{*}{45}  & \multirow{2}{*}{568} & \multirow{2}{*}{33102} & \multirow{2}{*}{23654} & \multirow{2}{*}{1.86} & 123.40 \\
    \cline{4-6} \cline{12-12}
            &   &   & Demod2 & 10.85  & 228.63 & & & & & & 122.92 \\
    \hline
    \cite{ref37} & 166 & 8 & - & 355.5 & 10.08 & 142.5 & 19 & 27506 & 29124 & - & - \\
    \hline
    \end{tabular}
    \label{tab:cascade}
\end{table*}

\subsection{Evaluation of Deployment Model}
We aim to support multiple NN variants on a single DDNA while preserving acceleration. Using the OFDM frame in Table~\ref{tab:ofdm} and QPSK, we train a cascaded model (DFT-Net $\rightarrow$ Demod-Net) for both DL-Transceiver and DL-Receiver; their hardware deployments are thus identical, differing only in parameters. The trained model is compiled with Vitis AI and mapped to the DPU, and the same model is deployed on DDNA instances of different sizes. Results appear in Table~\ref{tab:cascade}. For a fair baseline at 20~MHz sampling, we also implement conventional modules functionally equivalent to the NN blocks (DFT and demodulation) and compare with the $m$-QAM demodulator in~\cite{ref37}.

Only one model is mapped to the DPU due to unsatisfactory runtime. As a general-purpose accelerator, DPU resource usage reflects the need to support diverse models/operators, not a single mapped network, resulting in substantially higher utilization than the task-specific streaming DDNA. The DPU’s per-frame latency ($\approx 35$~ms) is an order of magnitude larger than DDNA, in part because unsupported operators fall back to the CPU, causing serial execution and frequent data transfers. In~\cite{ref37}, all parameters are on-chip, incurring heavy memory usage; limited compute parallelism yields longer latency and constrained maximum frequency.

By contrast, DDNA is tailored to the models in Sec.~\ref{subsec:network}, enabling dataflow optimizations and on-PL placement of manipulation operators to reduce off-/on-chip movement (Fig.~\ref{fig_7}). DDNA achieves microsecond-scale latency (as low as $10.85~\mu$s), supporting sampling rates approaching 60~MHz. Throughput also improves when Demod-Net is added, nearly $10\%$ over DFT-Net alone on DDNA($32\times16$), due to full pipelining and temporal overlap of stages. The DPU exhibits the highest power consumption because generality increases control and memory overheads; with small operators and high bandwidth, array utilization is low and memory traffic dominates, degrading energy efficiency (GOPS/W). DDNA attains higher efficiency via on-chip reuse and minimal control. Although DDNA($32\times16$) doubles the throughput of DDNA($16\times16$), its power rises by more than $2\times$, reducing efficiency—likely due to superlinear growth in interconnect/control complexity and tool-induced overheads (synthesis/place–route).

Table~\ref{tab:para} details representative layer parameters and measured cycles on DDNA($16\times16$). For DFT-Net, Linear layers mapped to the PEA achieve cycles $\approx 1/256$ of MACs, matching the $16\times 16$ array and indicating full utilization. Because the two Linear layers are identical, either separate (Linear-1/2) or merged (Linear-12) execution can be used; merged execution yields fewer cycles than the sum due to the data-merging scheme in Sec.~\ref{subsec:merge}. In Demod-Net, a row-vector input activates only one PEA row, inflating cycles to $\approx 1/16$ of MACs. After applying data merging, 16 input groups are injected concurrently across rows, keeping cycles unchanged while restoring peak utilization—thus the latencies in Table~\ref{tab:cascade} should be interpreted as averages over multiple runs.

\begin{table}[t]
    \centering
    \renewcommand{\arraystretch}{1.5}
    \caption{Representative Layer Parameters and DDNA($16\times16$) Results}
    \resizebox{\columnwidth}{!}{%
    \begin{tabular}{|c|c|c|c|c|c|c|c|}
    \hline
    \multirow{2}{*}{\textbf{Model}} & \multirow{2}{*}{\textbf{Layer}} & \multicolumn{4}{c|}{\textbf{Parameters}} & \multirow{2}{*}{\textbf{MAC}} & \multirow{2}{*}{\textbf{Cycle}} \\
    \cline{3-6}
        &   & I & X & O & Y &  & \\
    \hline
    \multirow{3}{*}{DFT} & Linear-1/2  & 16 & 80 & 16 & 80 & 102400 & 431    \\
    \cline{2-8}
                             & Linear-12  & 16 & 80 & 16 & 80 & 204800 & 831     \\
    \cline{2-8}
                             & Conv1D  & 4 & 80 & 2 & 80 & 5120 & 80         \\ 
    \hline
    \multirow{2}{*}{Demod1} & Linear  & 1 & 1280 & 1 & 1472 & 1.88 M & 132960 \\
    \cline{2-8}
                            & Conv1D  & 4 & 736 & 2 & 736 & 5888 & 736     \\
    \hline
    \multirow{2}{*}{Demod2} & Linear-1/2  & 1 & 640 & 1 & 736 & 471040 & 34480 \\
    \cline{2-8}
                            & Linear-12  & 1 & 640 & 1 & 736 & 942080 & 66480 \\ 
    \cline{2-8}
                            & Conv1D  & 4 & 736 & 2 & 736 & 5888 & 736 \\
    \hline
    \end{tabular}
    }
    \label{tab:para}
\end{table}

\subsection{Adaptive Model}
Practical wireless links adapt subcarrier counts and modulation orders to channel conditions. Motivated by this, we develop DL-based models and propose DDNA as an adaptive deployment substrate. Trained model parameters are precomputed and stored in DDR; when the modulation changes, the system adapts by loading new parameters without hardware modification. An auxiliary controller is required to detect operating-point changes. Although off-chip parameter storage introduces movement overhead, the pipelining strategy in Sec.~\ref{subsec:ps} effectively hides this latency.

Realistic channels introduce multipath, fading, and interference, necessitating channel estimation and equalization~\cite{ref17}. This work establishes a foundation under AWGN, demonstrates DL-driven gains, and completes hardware deployment. Future work will extend to richer channel models and leverage deep learning’s function-approximation capacity to minimize end-to-end loss. On the hardware side, we will further refine DDNA to support efficient deployment of a broader class of models.

\section{Conclusion}\label{section:Conclusion}
In this work, we proposed DFT-Net and Demod-Net to replace the IDFT/DFT and demodulation modules in an OFDM transceiver, and we examined their feasibility for mapping onto a SoC. Using extensive randomized tests across a range of SNR, the DL-based OFDM transceiver achieves performance that is comparable to, or better than, conventional counterparts under multiple modulation schemes. We also designed and implemented DDNA on an FPGA to deploy different models and accelerate execution. Compared with hardware realizations of conventional algorithms, DDNA sustains a 38–66\% reduction in latency while maintaining a 1.5–2 dB gain, and it consumes significantly fewer hardware resources than general-purpose accelerators. Future work will refine both model architectures and deployment strategies under dynamic wireless conditions to further improve the generalization and robustness of DL-enabled transceivers.




 
%

\newpage

 




\vfill

\end{document}